# Asexual and sexual replication in sporulating organisms

Bohyun Lee and Emmanuel Tannenbaum*

*School of Biology, Georgia Institute of Technology, Atlanta, GA 30332*

This paper develops models describing asexual and sexual replication in sporulating organisms. Replication via sporulation is the replication strategy for all multicellular life, and may even be observed in unicellular life (such as with budding yeast). We consider diploid populations replicating via one of two possible sporulation mechanisms: (1) Asexual sporulation, whereby adult organisms produce single-celled diploid spores that grow into adults themselves. (2) Sexual sporulation, whereby adult organisms produce single-celled diploid spores that divide into haploid gametes. The haploid gametes enter a haploid "pool", where they may recombine with other haploids to form a diploid spore that then grows into an adult. We consider a haploid fusion rate given by second-order reaction kinetics. We work with a simplified model where the diploid genome consists of only two chromosomes, each of which may be rendered defective with a single point mutation of the wild-type. We find that the asexual strategy is favored when the rate of spore production is high compared to the characteristic growth rate from a spore to a reproducing adult. Conversely, the sexual strategy is favored when the rate of spore production is low compared to the characteristic growth rate from a spore to a reproducing adult. As the characteristic growth time increases, or as the population density increases, the critical ratio of spore production rate to organism growth rate at which the asexual strategy overtakes the sexual one is pushed to higher values. Therefore, the results of this model suggest that, for complex multicellular organisms, sexual replication is favored at high population densities, and low growth and sporulation rates.

## I. INTRODUCTION

The emergence of sexual replication as the preferred replication strategy for complex multicellular organisms is one of the oldest and most important problems in evolutionary biology [1]. There are two broad theories for the selective advantage for sex [1–6], both of which have more than one specific version.

The first theory for the selective advantage for sex is the genetic repair theory, or, in the context of small populations, the Muller Ratchet theory [4, 6–10]. The genetic repair theory holds that sex evolved as a way to remove mutations from diploid genomes. By reproducing via a haploid intermediate, defective copies of genes can be discarded, and functional copies of genes can be brought together via haploid fusion.

In the context of finite populations, the genetic repair theory takes the form of the Muller Ratchet theory [7–9]. Briefly, Muller's Ratchet is a phenomenon whereby a finite population will steadily accumulate mutations, and may eventually go extinct as a result. Sexual replication, by providing a mechanism to discard defective genes, can slow down or stop the Muller's Ratchet, and thereby prevent small populations from going extinct.

The second theory for the selective advantage for sex is the adaptability theory [11]. This theory, which originates with Weismann, has two versions: The Vicar of Bray hypothesis, and the Red Queen hypothesis. The Vicar of Bray hypothesis argues that sex increases variability in small populations, making them more adaptable to changing circumstances. The name derives from an English cleric who was known for changing his religion according to circumstance [1]. Within the context of the idea that sex increases variation within a population, it is believed that host-parasite co-evolutionary dynamics drove the emergence of sex [6, 11]. This "arms race" argument for the existence of sex is termed the Red Queen hypothesis.

While the various theories for sex are not necessarily mutually contradictory, each of them is either incomplete or has difficulties. The Muller's Ratchet theory, for example, by relying on a small population size, suggests that sex should disappear in large populations. It is not immediately clear that this should be true, however, since many sexual organisms can attain seemingly fairly large population sizes.

The Vicar of Bray and the Red Queen Hypotheses rely on a dynamic environment. However, there are numerous sexually replicating organisms that have remained essentially unevolved over millions of years, in what appear to be fairly stable environments. Therefore, it is not immediately clear that a dynamic environment is the main selective pressure driving the emergence of sex.

The genetic repair theory is the theory with the broadest acceptance among evolutionary biologists. That being said, it could be argued that the theory is incomplete, because it is does not explain why sexual replication is disadvantageous in organisms such as bacteria [12]. Furthermore, it does not explain why, among the organisms that replicate sexually, some employ sexual replication merely as a stress response (such as Baker's yeast), and why others replicate exclusively via the sexual pathway.

It should be noted here that the issue is not why recombination is beneficial, since genetic recombination oc-curs at all biological length scales [13]. The issue is why some organisms, such as bacteria, replicate asexually (and sometimes exchange genetic material with another), and why other organisms, such as humans, replicate via a haploid intermediate.

In two recent papers [14, 15], Tannenbaum and Fontanari developed simple evolutionary dynamics models based on the sexual stress response in Baker's yeast. The first paper assumed a density-independent characteristic haploid fusion time, while the second assumed second-order haploid fusion kinetics. In both papers, the general result that emerged was that sex is favored when the characteristic haploid fusion time is small compared with the characteristic replication time. Therefore, the results of the models suggested that sex is favored in slowly replicating organisms, and that, all other factors being equal, sex is favored at high population densities.

While the results of these models were broadly consistent with what is observed biologically, they are unsuitable for considering additional features of sexual replication, such as gamete differentiation into sperm and egg, and sex differentiation into male and female. Although many organisms replicate sexually, not all have distinct gametes, and, even among organisms that produce distinct gametes (the anisogamous organisms), not all have distinct male and female sexes. There are therefore presumably different regimes where the different types of sexual strategies will be advantageous.

The previous two models, because they consider a sexual pathway whereby a given diploid cell splits into two haploids, do not reflect the fact that most sexually replicating organisms may continually produce gametes that may recombine to produce new organisms. The analogous process for asexually replicating organisms is sporulation. Therefore, before considering the regimes where asexual replication and various sexual replication strategies are advantageous, we first need to develop a formalism for describing the evolutionary dynamics of populations which replicate by producing either single-celled spores or gametes.

This paper is divided into four sections: In Section II, we develop a model describing asexual replication of organisms that replicate by producing single-celled spores. In Section III, we consider the analogous process for sexual replication, where we assume that the gametes fuse via second-order kinetics. In Section IV, we numerically solve for the steady-state mean fitnesses of the various populations, and determine the various regimes where the different strategies are advantageous. We find that sexual replication is favored in organisms that produce offspring at lower rates. The critical offspring production rate increases as the characteristic organismal maturation time increases, or as the density of adult organisms increases. Therefore, the results of this paper appear to be broadly consistent with what is observed biologically. However, based on known scaling laws for organismal maturation times as a function of organism size, it is not clear that the results of our model imply that sexual replication will be favored in larger organisms. In Section V, we summarize our conclusions, and discuss future research directions. In particular, we discuss additional modeling that could resolve this possible inconsistency between our predictions for the preferred sexual strategy as a function of organism size, and what is observed biologically.

We should note that the models we develop are quasispecies-type models. Quasispecies theory has found application in a wide range of problems in evolutionary dynamics, including molecular, viral, and bacterial evolution, the immune response, and the emergence of cancer [16–47]. The formalism is well-suited to studying the selective advantage for various types of replication strategies.

## II. ASEXUAL REPLICATION VIA SPORULATION

In this section, we develop the evolutionary dynamics equations appropriate for describing asexual replication via sporulation. As discussed in the Introduction, while a model involving simple binary fission is appropriate for unicellular organisms, such an approach is inadequate for describing larger, multicellular organisms. The reason for this is that asexually replicating larger organisms begin their life cycle as a single cell. This cell develops into the adult organism, which then begins to produce single-celled spores at some rate (see Figure 1).

Therefore, in order to properly study the regimes where asexual and various sexual replication strategies are advantageous, we first need to develop equations more appropriate for the replication dynamics of multicelled organisms.

### A. Definitions

The asexual replication of an organism (single or multi-celled) by sporulation occurs as follows: An immature organism grows to adult size. The adult then produces single-celled spores by budding. Each of these spores are immature organisms that then repeat the process and develop into adults on their own.

For simplicity, we assume that the organisms have diploid genomes consisting of two chromosomes. The genome of each organism may then be denoted by $\{\sigma_1, \sigma_2\}$, where $\sigma_1$ and $\sigma_2$ denote the respective base sequences of each chromosome (we assume that each chromosome consists of a single strand of bases, so that a given chromosome $\sigma$ may be written as $\sigma = s_1 s_2 \ldots s_L$. Here, $L$ is the total sequence length, and each $s_i$ is a base which is chosen from an alphabet of size $S$, where $S = 4$ for terrestrial life).

Following standard practice in quasispecies theory, we make the simplifying assumption that there exists a single master sequence, denoted $\sigma_0$, for which a given chro-mosome is functional. In this single-fitness-peak approximation, any chromosome $\sigma \neq \sigma_0$ is non-functional. A given genome may therefore be classified by the number of functional chromosomes it has, namely zero, one, or two.

A genome with zero functional chromosomes is said to be of type $(u, u)$, where $u$ signifies that a chromosome is *unviable*. A genome with one functional chromosome is said to be of type $(v, u)$, where $v$ signifies that a chromosome is *viable*. A genome with two functional chromosomes is said to be of type $(v, v)$.

We assume that the growth of an immature organism to adulthood is characterized by a first-order rate constant. We further assume that this rate constant is genome-dependent, since different organisms are differently suited to the given environment, and so will reach maturity at different rates. We therefore let $\kappa_{vv}$, $\kappa_{vu}$, and $\kappa_{uu}$ denote the first-order spore-to-adult rate constants for the $(v,v)$, $(v,u)$, and $(u,u)$ organisms, respectively.

We also assume that, once an organism reaches adulthood, it produces spores at some fixed rate, which is again genome-dependent. The rates of per-organism spore production for the various genome types are denoted $\omega_{vv}$, $\omega_{vu}$, and $\omega_{uu}$.

We let $n_{ai,xy}$ denote the number of asexual, immature organisms with genome $(x, y)$, and $n_{am,xy}$ denote the number of asexual, mature organisms with genome of type $(x, y)$.

During the sporulation process itself, the adult organism replicates each of the chromosomes in the genome. It is assumed that the adult organism retains the original parent chromosomes, and only the two newly synthesized daughters segregate into the budding spore. This "immortal strand" mechanism is believed to occur in the adult stem cells of vertebrates and in budding yeast [48–50]. Since the spores are being produced by stem cells in the adult, it is reasonable to assume that a similar mechanism is at work here as well.

Finally, the replication of each chromosome is not error-free. We let $p$ denote the replication fidelity, defined as the probability that a $v$ chromosome produces a $v$ daughter. If we neglect backmutations, then a $v$ chromosome produces a $u$ daughter with probability $1-p$, and a $u$ chromosome produces a $u$ daughter with probability 1.

### B. Mutation-selection equations

The mutation-selection equations governing the evolutionary dynamics of the sporulating population are then given by,

$$\frac{dn_{am,vv}}{dt} = \kappa_{vv} n_{ai,vv}$$
$$\frac{dn_{am,vu}}{dt} = \kappa_{vu} n_{ai,vu}$$
$$\frac{dn_{am,uu}}{dt} = \kappa_{uu} n_{ai,uu}$$

$$\frac{dn_{ai,vv}}{dt} = p^2 \omega_{vv} n_{am,vv} - \kappa_{vv} n_{ai,vv}$$
$$\frac{dn_{ai,vu}}{dt} = 2p(1-p)\omega_{vv} n_{am,vv} + p\omega_{vu} n_{am,vu} - \kappa_{vu} n_{ai,vu}$$
$$\frac{dn_{ai,uu}}{dt} = (1-p)^2 \omega_{vv} n_{am,vv} + (1-p)\omega_{vu} n_{am,vu}$$
$$+ \omega_{uu} n_{am,uu} - \kappa_{uu} n_{ai,uu} \quad (1)$$

We now define $n_{am} = n_{am,vv} + n_{am,vu} + n_{am,uu}$, so that $n_{am}$ is simply the total population of mature adults. We then define all population fractions with respect to this population number. Specifically, we define the population fractions $x_{aq,rs} = n_{aq,rs}/n_{am}$, where $q = m, i$, and $(r, s)$ is the genome type. Note that, in principle, some of the population fractions can be greater than 1, since they are not defined with respect to the total population, but rather with respect to the population of mature adults.

Changing variables, we obtain,

$$\frac{dx_{am,vv}}{dt} = \kappa_{vv} x_{ai,vv} - \bar{\kappa}_a(t) x_{am,vv}$$
$$\frac{dx_{am,vu}}{dt} = \kappa_{vu} x_{ai,vu} - \bar{\kappa}_a(t) x_{am,vu}$$
$$\frac{dx_{am,uu}}{dt} = \kappa_{uu} x_{ai,uu} - \bar{\kappa}_a(t) x_{am,uu}$$
$$\frac{dx_{ai,vv}}{dt} = p^2 \omega_{vv} x_{am,vv} - (\kappa_{vv} + \bar{\kappa}_a(t)) x_{ai,vv}$$
$$\frac{dx_{ai,vu}}{dt} = 2p(1-p)\omega_{vv} x_{am,vv} + p\omega_{vu} x_{am,vu}$$
$$- (\kappa_{vu} + \bar{\kappa}_a(t)) x_{ai,vu}$$
$$\frac{dx_{ai,uu}}{dt} = (1-p)^2 \omega_{vv} x_{am,vv} + (1-p)\omega_{vu} x_{am,vu} + \omega_{uu} x_{am,uu}$$
$$- (\kappa_{uu} + \bar{\kappa}_a(t)) x_{ai,uu} \quad (2)$$

where $\bar{\kappa}_a(t) = \kappa_{vv} x_{ai,vv} + \kappa_{vu} x_{ai,vu} + \kappa_{uu} x_{ai,uu}$ is simply the mean fitness of the population.

### C. Steady-state mean fitness

We may solve for the steady-state of this system of equations analytically. To begin, we first assume that $\kappa_{uu} = 0$, so that, the first three equations give, at steady-state, that,

$$x_{ai,vv} = \frac{\bar{\kappa}_a(t=\infty)}{\kappa_{vv}} x_{am,vv}$$
$$x_{ai,vu} = \frac{\bar{\kappa}_a(t=\infty)}{\kappa_{vu}} x_{am,vu}$$
$$x_{am,uu} = 0 \quad (3)$$

so that,

$$0 = [p^2 \kappa_{vv} \omega_{vv} - (\kappa_{vv} + \bar{\kappa}_a(t=\infty))\bar{\kappa}_a(t=\infty)] x_{am,vv} \quad (4)$$

If $x_{am,vv} > 0$, then we have,

$$\bar{\kappa}_a(t=\infty)^2 + \kappa_{vv} \bar{\kappa}_a(t=\infty) - p^2 \kappa_{vv} \omega_{vv} = 0 \quad (5)$$

so that,

$$\bar{\kappa}_a(t=\infty) = \bar{\kappa}_{a,1}(t=\infty) = \frac{1}{2}\kappa_{vv}[-1 + \sqrt{1 + 4p^2\frac{\omega_{vv}}{\kappa_{vv}}}] \quad (6)$$

If $x_{am,vv} = 0$, then we have,

$$0 = [p\kappa_{vu}\omega_{vu} - (\kappa_{vu} + \bar{\kappa}_a(t=\infty))\bar{\kappa}_a(t=\infty)]x_{am,vu} \quad (7)$$

so, if $x_{am,vu} > 0$, then we have,

$$\bar{\kappa}_a(t=\infty) = \bar{\kappa}_{a,2}(t=\infty) = \frac{1}{2}\kappa_{vu}[-1 + \sqrt{1 + 4p\frac{\omega_{vu}}{\kappa_{vu}}}] \quad (8)$$

Based on a stability analysis of the possible steady-states, it is possible to show that the actual value of $\bar{\kappa}_a(t=\infty)$ is given by $\max\{\kappa_{a,1}(t=\infty), \kappa_{a,2}(t=\infty)\}$.

Assuming that $\kappa_{vv} > \kappa_{vu}$, and that $\omega_{vv}/\kappa_{vv} = \omega_{vu}/\kappa_{vu}$, we have that $\bar{\kappa}(t=\infty) = \bar{\kappa}_{a,1}$ at $p=1$. Defining $\alpha = \kappa_{vu}/\kappa_{vv}$ and $\beta = \omega_{vv}/\kappa_{vv}$, we obtain an error threshold transition at some $p_{crit} = p_{crit}(\alpha, \beta)$, defined by the condition $\bar{\kappa}_{a,1}(t=\infty) = \bar{\kappa}_{a,2}(t=\infty)$. Below this replication fidelity, the effective growth rate of the $(v, v)$ genomes is no longer competitive with that of the $(v, u)$ genomes, and the result is the disappearance of the $(v, v)$ genomes from the population.

## III. SEXUAL REPLICATION

In this section we consider a sexual replication pathway, whereby a given diploid spore splits into two haploids, which may then fuse with other haploids in the population.

### A. Definitions

In the sexual replication model being considered here, a mature diploid with genome $(v, v)$ produces smaller diploid spores at a rate given by $\omega_{vv}$. These diploid spores then divide into two haploid intermediates. The haploids of type $v$ may then recombine with each other to form an immature diploid of type $(v, v)$, which then grows to a mature diploid and begins the cycle again.

In this model, we assume that only the $v$ haploids may recombine with one another. Essentially, the $u$ haploids are defective and cannot participate further in the replication process. We assume that the haploid fusion rate is described by second-order kinetics characterized by a rate constant $\gamma$. We also assume that the haploids have a finite lifetime in the population, and decay with a first-order rate constant $\kappa_h$.

Because only viable haploids can recombine, the only diploids in the population are immature and mature diploids of type $(v, v)$. We let $n_{si,vv}$, $n_{sm,vv}$ denote the number of immature and mature diploids of type $(v, v)$, respectively. We also let $n_v$ denote the number of viable haploids in the population.

### B. Mutation-selection equations

For a haploid fusion rate governed by second-order reaction kinetics, the equations governing the evolutionary dynamics of the population are given by,

$$\frac{dn_{sm,vv}}{dt} = \kappa_{vv} n_{si,vv}$$
$$\frac{dn_{si,vv}}{dt} = \frac{1}{2}\frac{\gamma}{V}n_v^2 - \kappa_{vv} n_{si,vv}$$
$$\frac{dn_v}{dt} = 2\omega_{vv} p n_{sm,vv} - \frac{\gamma}{V}n_v^2 - \kappa_h n_v \quad (9)$$

where $V$ denotes the system volume in which the organisms are present. In this model, we assume that the system volume changes in such a way as to maintain a constant density $\rho$ of mature diploids (they are the fully grown organisms, so the total volume is dictated by the number of mature diploids present). We also define all population fractions with respect to the mature diploids, so that $x_{si,vv} \equiv n_{si,vv}/n_{sm,vv}$, and $x_v \equiv n_v/n_{sm,vv}$.

In terms of the population fractions, we have,

$$\frac{dx_{si,vv}}{dt} = \frac{1}{2}\gamma\rho x_v^2 - (\kappa_{vv} + \bar{\kappa}_s(t))x_{si,vv}$$
$$\frac{dx_v}{dt} = 2\omega_{vv} p - \gamma\rho x_v^2 - (\kappa_h + \bar{\kappa}_s(t))x_v \quad (10)$$

where the mean fitness $\bar{\kappa}_s(t) = (1/n_{sm,vv})dn_{sm,vv}/dt = \kappa_{vv} x_{si,vv}$. Note that we do not need to include an equation for $x_{sm,vv}$, since by definition $x_{sm,vv} = 1$.

### C. Steady-state mean fitness

Using the relationship between $\bar{\kappa}_s(t)$ and $x_{si,vv}$, we obtain the steady-state equations,

$$0 = \frac{1}{2}\gamma\rho x_v^2 - (1 + \frac{\bar{\kappa}_s(t=\infty)}{\kappa_{vv}})\bar{\kappa}_s(t=\infty)$$
$$0 = 2\omega_{vv} p - \gamma\rho x_v^2 - (\kappa_h + \bar{\kappa}_s(t=\infty))x_v \quad (11)$$

The first equation may be solved for $x_v$, giving,

$$x_v = \sqrt{\frac{2(1 + \frac{\bar{\kappa}_s(t=\infty)}{\kappa_{vv}})\frac{\bar{\kappa}_s(t=\infty)}{\kappa_{vv}}}{\frac{\gamma\rho}{\kappa_{vv}}}} \quad (12)$$

Plugging into the second equation, and re-arranging, we obtain,

$$\frac{1}{2}\frac{\kappa_{vv}}{\gamma\rho} = \frac{(\frac{\omega_{vv}}{\kappa_{vv}}p - (1+\frac{\bar{\kappa}_s(t=\infty)}{\kappa_{vv}})\frac{\bar{\kappa}_s(t=\infty)}{\kappa_{vv}})^2}{(\frac{\kappa_h}{\kappa_{vv}} + \frac{\bar{\kappa}_s(t=\infty)}{\kappa_{vv}})^2(1+\frac{\bar{\kappa}_s(t=\infty)}{\kappa_{vv}})\frac{\bar{\kappa}_s(t=\infty)}{\kappa_{vv}}} \quad (13)$$

If we assume that $\kappa_{vv}/\gamma\rho \to 0$, then we obtain,

$$0 = (\frac{\bar{\kappa}_s(t=\infty)}{\kappa_{vv}})^2 + \frac{\bar{\kappa}_s(t=\infty)}{\kappa_{vv}} - p\frac{\omega_{vv}}{\kappa_{vv}} \quad (14)$$

which may be solved to give,

$$\bar{\kappa}_s(t=\infty) = \frac{1}{2}\kappa_{vv}[-1+\sqrt{1+4p\frac{\omega_{vv}}{\kappa_{vv}}}] \qquad (15)$$

Note that, at least in the limit of $\kappa_{vv}/\gamma\rho \to 0$, we have $\bar{\kappa}_s(t=\infty) \geq \bar{\kappa}_a(t=\infty)$, with equality only occurring at $p = 0, 1$, or if $\kappa_{vv} = \kappa_{vu}$. Therefore, in the model being considered here, sexual replication outcompetes asexual replication when the cost for sex (as measured by the ratio of the characteristic haploid fusion time to the characteristic growth time) is negligible.

## IV. COMPARISON OF THE VARIOUS REPLICATION MECHANISMS

We may numerically solve for the mean fitness of the sexual replication mechanism, and compare the values of $\bar{\kappa}_a(t=\infty)/\kappa_{vv}$ and $\bar{\kappa}_s(t=\infty)/\kappa_{vv}$. The population with the larger of the two values will be selected for in the given parameter regime. For simplicity, we assume that $\kappa_h = 0$. In this case, the steady-state mean fitness of the sexual population is obtained by solving,

$$\frac{1}{2}\frac{\kappa_{vv}}{\gamma\rho} = \frac{(\frac{\omega_{vv}}{\kappa_{vv}}p - (1+\frac{\bar{\kappa}_s(t=\infty)}{\kappa_{vv}})\frac{\bar{\kappa}_s(t=\infty)}{\kappa_{vv}})^2}{(\frac{\bar{\kappa}_s(t=\infty)}{\kappa_{vv}})^3(1+\frac{\bar{\kappa}_s(t=\infty)}{\kappa_{vv}})} \qquad (16)$$

so that $\bar{\kappa}_s(t=\infty)/\kappa_{vv}$ depends only on $\kappa_{vv}/\gamma\rho$ and $\omega_{vv}/\kappa_{vv}$ for a given value of $p$.

The asexual mean fitness is given by $\bar{\kappa}_a(t=\infty)/\kappa_{vv} = \max\{\bar{\kappa}(t=\infty)_{a,1}/\kappa_{vv}, \bar{\kappa}(t=\infty)_{a,2}/\kappa_{vv}\}$, where,

$$\frac{\bar{\kappa}_{a,1}(t=\infty)}{\kappa_{vv}} = \frac{1}{2}[-1+\sqrt{1+4\frac{\omega_{vv}}{\kappa_{vv}}p^2}]$$
$$\frac{\bar{\kappa}_{a,2}(t=\infty)}{\kappa_{vv}} = \frac{1}{2}\alpha[-1+\sqrt{1+4\frac{\omega_{vv}}{\kappa_{vv}}p}] \qquad (17)$$

where we have assumed that $\omega_{vu}/\kappa_{vu} = \omega_{vv}/\kappa_{vv}$.

For a given $p$ we can determine, as a function of $\omega_{vv}/\kappa_{vv}$, the value of $\kappa_{vv}/\gamma\rho$ at which the sexual and asexual mean fitnesses are equal. This curve defines the boundary separating the regimes of dominance for the two strategies.

We can analytically determine the behavior of this curve in both the low and high $\omega_{vv}/\kappa_{vv}$ regimes. First, when $\omega_{vv}/\kappa_{vv}$ is small, then we have,

$$\frac{\bar{\kappa}_{a,1}(t=\infty)}{\kappa_{vv}} = \frac{\omega_{vv}}{\kappa_{vv}}p^2$$
$$\frac{\bar{\kappa}_{a,2}(t=\infty)}{\kappa_{vv}} = \frac{\omega_{vv}}{\kappa_{vv}}\alpha p \qquad (18)$$

and so, in any event, we have $\bar{\kappa}_a(t=\infty)/\kappa_{vv} = (\omega_{vv}/\kappa_{vv})\chi p$, where $\chi = \alpha, p$.

Now, since we wish to determine the value of $\kappa_{vv}/\gamma\rho$ at which the sexual and asexual mean fitnesses are equal, we simply use the value of $\bar{\kappa}_a(t=\infty)/\kappa_{vv}$ for $\bar{\kappa}_s(t=\infty)/\kappa_{vv}$ in Eq. (16). Since we are assuming $\omega_{vv}/\kappa_{vv}$ is small, we can drop all higher-order terms in the numerator and denominator expressions, giving,

$$\frac{\kappa_{vv}}{\gamma\rho} = 2\frac{(1-\chi)^2}{\chi^3 p}(\frac{\omega_{vv}}{\kappa_{vv}})^{-1} \qquad (19)$$

Now, when $\omega_{vv}/\kappa_{vv}$ is large, then we have,

$$\frac{\bar{\kappa}_{a,1}(t=\infty)}{\kappa_{vv}} = p\sqrt{\frac{\omega_{vv}}{\kappa_{vv}}}$$
$$\frac{\bar{\kappa}_{a,2}(t=\infty)}{\kappa_{vv}} = \alpha p^{1/2}\sqrt{\frac{\omega_{vv}}{\kappa_{vv}}} \qquad (20)$$

and so, in any event, we have $\bar{\kappa}_a(t=\infty)/\kappa_{vv} = p^{1/2}\chi(\omega_{vv}/\kappa_{vv})^{1/2}$, where $\chi = p^{1/2}, \alpha$. Following a similar procedure to the one used for small $\omega_{vv}/\kappa_{vv}$ (only this time we drop the lowest-order terms since $\omega_{vv}/\kappa_{vv}$ is large), we obtain that the asexual and sexual mean fitnesses are equal when,

$$\frac{\kappa_{vv}}{\gamma\rho} = 2(\frac{1}{\chi^2}-1)^2 \qquad (21)$$

We therefore find that, at low sporulation rates, the cost for sex, as measured by $\kappa_{vv}/\gamma\rho$, must be made steadily larger as $\omega_{vv}/\kappa_{vv}$ decreases in order for the asexual strategy to remain competitive with the sexual one. We find that this pattern indeed holds at all values of $\omega_{vv}/\kappa_{vv}$. The large $\omega_{vv}/\kappa_{vv}$ behavior of this model also shows that, once the cost for sex drops below a critical value (given by $2(1/\chi^2 - 1)^2$), then sexual replication is the dominant strategy for all values of $\omega_{vv}/\kappa_{vv}$.

The cost for sex is measured by $\kappa_{vv}/\gamma\rho$, because this quantity measures the ratio of the characteristic haploid fusion time ($\propto 1/\gamma\rho$) to the characteristic maturation time ($\propto 1/\kappa_{vv}$). What the results of this model suggest is that, as the sporulation rate decreases (as measured by $\omega_{vv}/\kappa_{vv}$), the critical cost for sex at which asexual replication overtakes sexual replication increases as well. If the cost for sex is sufficiently small, however, ($< 2(1/\chi^2 - 1)^2$), then sexual replication outcompetes asexual replication for all sporulation rates.

These results may be understood as follows: As the sporulation rate drops, the time interval between the production of diploid spores increases. Therefore, the additional fitness penalty incurred by having the diploid spores split and pay a time cost in finding other haploids with which to fuse decreases. The result is that the cost for sex has to be pushed to higher values before the strategy becomes disadvantageous.

It is also interesting to note from the small $\omega_{vv}/\kappa_{vv}$ expression for the critical value of $\kappa_{vv}/\gamma\rho$ that, as $p$ decreases, the critical cost for sex increases. This makes sense since, as the replication fidelity drops, the benefit of sexual recombination increases as well.

Therefore, the results of this model suggest that, in the case of multicellular organisms that replicate via the production of unicellular spores (or gametes, in the case of sexual organisms), sexual replication is favored in organisms that sporulate slowly. This is essentially equivalent to the statement that sexual replication is favored in organisms that produce few offspring. For such organisms, the time cost associated with haploid fusion is comparatively small, and so it makes sense to replicate via a mechanism that provides the few offspring produced with the highest possible survival probability.

Furthermore, as the maturation time of the organisms increases, or, as the population density increases, the sexual strategy becomes advantageous as well. Again, this makes sense, since both an increased maturation time and an increased population density reduce the time cost associated with sex, thereby leading to a selective advantage for the sexual replication strategy.

## V. CONCLUSIONS AND FUTURE RESEARCH

This paper developed models describing asexual and sexual replication in sporulating organisms. Such models are necessary for studying the selective advantages of asexual and sexual replication strategies in multicellular organisms.

Under the assumption of second-order haploid fusion kinetics, we found that sexual replication is favored at low sporulation rates, or equivalently, at low offspring numbers, long organism maturation times, and high population densities. These results make sense, since low offspring numbers, long organism maturation times, and high population densities all reduce the effective time cost associated with sex. In particular, for organisms that produce relatively few offspring, it makes sense to invest a comparatively small cost in sex and thereby maximize the survival probability of each child.

While the results of this paper appear to be consistent with actual behavior, a difficulty that arises is in correlating preferred replication strategy with organism size. It is well-known that the maturation time of an organism goes as $N^{1/4}$, where $N$ is the number of cells in the adult organism [51, 52]. Therefore, as a function of $N$, we expect $\kappa_{vv}$ to scale as $N^{-1/4}$. The population density of adults, however, should scale as $1/N$, giving $\kappa_{vv}/\gamma\rho \propto N^{3/4}$. Therefore, the cost of sex goes to $\infty$ as $N \to \infty$, and so, from the small $\omega_{vv}/\kappa_{vv}$ criterion, we obtain that $\omega_{vv}/\kappa_{vv} \propto N^{-3/4}$ for large $N$, so that $\omega_{vv} \propto 1/N$. Therefore, as organism size increases, the rate of sporulation must scale as the reciprocal of the organism size. This rate of decrease appears to be much too rapid to allow for the predominance of sex as the preferred replication strategy of larger organisms.

To hopefully resolve this issue, future research will study the effects of mobile gametes on haploid fusion rates. In this vein, gamete differentiation (sperm/egg) and sex differentiation (male/female) will be studied as well. It is possible that, as organism size increases, these replication strategies reduce the time cost for sex to a sufficient amount that sex does indeed emerge as the preferred replication strategy for larger organisms. While important work on gamete differentiation has been done by Dusenbery [53–55], a determination of the regimes where the various sexual replication strategies and asexual replication are advantageous has not yet been done.

Another important factor that will need to be considered is death. Thus far, our models do not assume that organisms eventually die. The neglect of this phenomenon could lead to an unrealistically large regime where asexual replication is dominant, though it is of course not yet clear if this is the case.

The dependence of $\alpha$ and $p$ on organism size and complexity will need to be considered as well. As organismal complexity grows, the size of the genome grows, and so it is likely that $\alpha$ and $p$ decrease. The result is that $\omega_{vv}/\kappa_{vv}$ must be pushed to higher values before asexual replication becomes the dominant replication strategy, and this increase in the critical sporulation rate might be sufficient to result in sexual replication being the dominant replication strategy at larger organism sizes.

Finally, a key assumption of our model is that the sexual organisms release gametes continuously. In reality, sexually replicating organisms generally store up gametes for a certain period, and then, during a mating season, collectively release these gametes into the surroundings (we are considering organisms that replicate in aqueous environments for this analysis). Thus, although the gamete production rate may be low, during these brief periods when massive numbers of gametes are rapidly released, the rate of haploid fusion is fast, thereby reducing the time cost associated with sex. This reduction in time cost could be quite large, which could significantly increase the size of the regime where the sexual strategy is dominant. In this case, a weaker dependence of $\omega_{vv}$ on $N$ may be necessary to ensure the selection for a sexual strategy, so that we do indeed obtain that sexual replication is preferred as organism size grows.

### Acknowledgments

The authors thank Jose F. Fontanari for helpful discussions leading to the completion of this work.